\documentclass[reprint,superscriptaddress,amsmath,amssymb,aps,floatfix]{revtex4-1}
\usepackage{graphicx}
\usepackage{dcolumn}
\usepackage{upgreek}
\usepackage{gensymb}
\usepackage{float}
\usepackage{bm}

\begin{document}


\title{Current-induced modulation of interfacial Dzyaloshinskii-Moriya interaction}


\author{Naoaki Kato}
\affiliation{Department of Physics, The University of Tokyo, Tokyo 113-0033, Japan}

\author{Masashi Kawaguchi}
\affiliation{Department of Physics, The University of Tokyo, Tokyo 113-0033, Japan}

\author{Yong-Chang Lau}
\affiliation{Department of Physics, The University of Tokyo, Tokyo 113-0033, Japan}
\affiliation{National Institute for Materials Science, Tsukuba 305-0047, Japan}

\author{Toru Kikuchi}
\affiliation{Yukawa Institute for Theoretical Physics, Kyoto University, Kyoto 606-8502, Japan}

\author{Yoshinobu Nakatani}
\affiliation{University of Electro-Communications, Chofu 182-8585, Japan}

\author{Masamitsu Hayashi}
\email[]{hayashi@phys.s.u-tokyo.ac.jp}
\affiliation{Department of Physics, The University of Tokyo, Tokyo 113-0033, Japan}
\affiliation{National Institute for Materials Science, Tsukuba 305-0047, Japan}


\newif\iffigure
\figurefalse
\figuretrue

\date{\today}

\begin{abstract}
The Dzyaloshinskii-Moriya (DM) interaction is an antisymmetric exchange interaction that is responsible for the emergence of chiral magnetism. The origin of the DM interaction, however, remains to be identified albeit the large number of studies reported on related effects. It has been recently suggested that the DM interaction is equivalent to an equilibrium spin current density originating from spin-orbit coupling, an effect referred to as the spin Doppler effect. The model predicts that the DM interaction can be controlled by spin current injected externally. Here we show that the DM exchange constant ($D$) in W/CoFeB based heterostructures can be modulated with external current passed along the film plane. At higher current, $D$ decreases with increasing current, which we infer is partly due to the adiabatic spin transfer torque.
At lower current, $D$ increases linearly with current regardless of the polarity of current flow. 
The rate of increase in $D$ with the current density agrees with that predicted by the model based on the spin Doppler effect. 
These results imply that the DM interaction at the HM/FM interface partly originates from an equilibrium interface spin (polarized) current which can be modulated externally.
\end{abstract}

\pacs{}

\maketitle

Surface and interface effects play an increasingly dominant role in thin film heterostructures with large spin orbit coupling\cite{manchon2015review}. In heterostructures with ultrathin ferromagnetic layers, perpendicular magnetic anisotropy (PMA) emerges owing to the modification of the interface electronic structure\cite{johnson1995pma,ikeda2010pmamtj}. 
The magnetic exchange interaction can also be modified at interfaces\cite{fert1990interface,bode2017dmi,fert2013skyrmion}: when a magnetic layer is placed next to a non-magnetic layer with strong spin-orbit interaction, the Dzyaloshinskii-Moriya (DM) interaction\cite{dzyaloshinskii1958jetp,moriya1960anisotropic} appears and influences the ordering of the magnetic moments. The DM interaction can stabilize a homochiral N\'{e}el domain wall\cite{heide2008dmi,ryu2013chiral,emori2013dmi,chen2013chiral1,tetienne2015dmi} in systems that will otherwise favor a non-chiral Bloch domain wall\cite{koyama2011intrinsic}. 

Recent experiments have shown that chiral N\'{e}el walls can be driven by current\cite{ryu2013chiral,emori2013dmi,torrejon2014interface,yang2015domain}. When current is passed along thin film heterostructures that include a ferromagnetic metal (FM) layer and a heavy metal (HM) layer, the spin Hall effect (SHE)\cite{liu2012she,sinova2015spinhall} of the HM layer generates spin current that diffuses into the FM layer and exerts spin torque on the magnetic moments\cite{kim2013torque,garello2013torque}, resulting in motion of domain walls. The efficiency of such current induced motion of chiral N\'{e}el walls is determined by the strength of the DM interaction as well as the current-spin conversion efficiency, often parameterized by the spin Hall angle of the HM layer. 

Identifying systems with large DM interaction is one of the focuses in the field of spin orbitronics.
Experimentally, it has been shown that the material used for the HM layer defines the sign and strength of the DM interaction\cite{chen2013chiral2,torrejon2014interface,ryu2014chiral,hrabec2014dmi}. Microscopically, a number of models have been proposed to describe its origin\cite{yang2015dmiorigin,kim2013rashbadmi,freimuth2014dmi,belabbes2016hund,kikuchi2016dzyaloshinskii}. Recent reports suggest that the DM interaction is equivalent to an equilibrium spin current density originating from spin-orbit coupling\cite{kim2013rashbadmi,kikuchi2016dzyaloshinskii,freimuth2017dmi,tatara2019physE}. Such model, referred to as the spin Doppler effect, suggests that the DM interaction is not a given material parameter and that non-equilibrium spin current injected externally to the system can modify its magnitude. 

Here we show that the interfacial DM interaction in W/CoFeB heterostructures can be modulated by current. Using current induced motion of domain walls, we find that the DM exchange constant increases with increasing current at low current but drops at larger current. The current flow direction plays little role in setting the DM interaction. The rate at which DM exchange constant increases with increasing current density is close to what the spin Doppler effect predicts. 

Radio-frequency sputtering is used to deposit thin films on thermally oxidized silicon substrates\footnote{See supplemental Material for supporting experimental results, details of the micromagnetic simulations, and discussion on the spin Doppler effect. which includes 
Refs. \cite{rohart2013skyrmion,edelstein1990,zhang2004beta}}. We show representative results from a film structure that consists of substrate/Ta($d_{\mathrm{Ta}}$)/W(1)/Co$_{20}$Fe$_{60}$B$_{20}$(1)/MgO(2)/Ta(1) (thickness in nm) with $d_{\mathrm{Ta}}\sim 0.5$ nm (referred to as sample A) and $d_{\mathrm{Ta}}\sim 2.3$ nm (sample B). 
The two samples exhibit different  DM exchange interaction. Films with different stacking are also studied: details of the films are introduced appropriately.  All films are patterned to micron-sized wires using optical lithography and Ar ion milling. The width and length of the wires are typically $\sim 5 \ \upmu$m and $\sim 30 \ \upmu$m, respectively. 10 Ta/100 Au electrodes are made using optical lithography and lift-off processes. 
The saturation magnetization per unit volume $M_\mathrm{s}$ and the effective magnetic anisotropy energy density $K_\mathrm{eff}$ are measured using vibrating sample magnetometry (VSM): results are summarized in Table \ref{para_table}.

\begin{table}[t]
\caption{Material parameters of the samples studied. \label{para_table}}
\begin{ruledtabular}
	\begin{tabular}{cccc}
	Samples	&	$M_\mathrm{s}$  (kA/m)	&	$K_\mathrm{eff}$ (10$^5$ J/m$^3$)	&	$D$ (mJ/m$^2$)\footnote{ $D$ estimated from $v$ vs. pulse amplitude using Eq. (\ref{DMI_saturation_eq}) with current independent $D$. The fitting results are shown by the solid lines in Figs. \ref{DMI_saturation}(a,b) and \ref{DMI_voltage2}(a-c). The sign of $D$ is determined by the sign of $H_{\textrm{DM}}$ obtained for a given wall type (e.g. up-down wall).}	\\
	\hline
	A	&	930	&	4.0	&	0.24	\\
	B	&	1090	&	6.2	&	0.33 \\
	C	&	1250	&	3.8	&	0.16 \\
	E	&	950	&	4.9	&	0.19 \\
	F	&	1460	&	8.5	&	-0.39 \\
	\end{tabular}
\end{ruledtabular}
\end{table}

\begin{figure}[b]
	\includegraphics[scale=0.7]{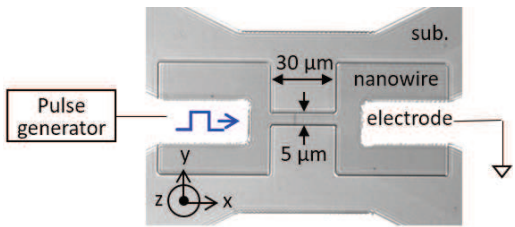}
	\caption{
	Kerr microscopy image of a representative wire used in the experiments. Definition of the coordinate axis and the measurement setup is illustrated. Positive current corresponds to current flow to $+x$ direction. \label{setup}}
\end{figure}

\begin{figure}[b]
\includegraphics[scale=0.45]{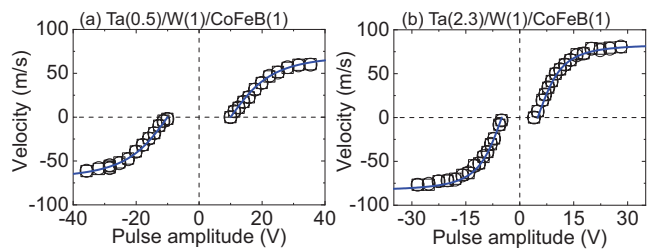}
	\caption{
	(a,b) Pulse amplitude dependence of domain wall velocity of (a) sample A and (b) sample B. The circles (squares) show velocity of up/down (down/up) domain walls. The blue solid lines show the fitting results using Eq. (\ref{DMI_saturation_eq}). The fitting parameters are summarized in Table~\ref{fit_table} of the Supplementary material. See supplementary material for the fitting of the pulse amplitude dependence of the velocity assuming a current dependent $D$ (Fig.~\ref{DMI_saturation_supp}).
\label{DMI_saturation}} 
\end{figure}

The motion of domain walls is studied using magneto-optical Kerr microscopy (Fig. \ref{setup}). The velocity of a domain wall is estimated by dividing the distance the wall traveled by the applied current pulse length. The current driven domain wall velocity of samples A and B is shown in Fig. \ref{DMI_saturation}(a) and \ref{DMI_saturation}(b) as a function of the applied pulse amplitude $I$. The pulse length is $\sim 9$ ns, which is made sufficiently short to avoid causing significant domain wall tilting\cite{ryu2012tilt,boulle2013tilt,safeer2016tilt}. The velocity saturates as the pulse amplitude increases, a characteristic often observed for spin Hall driven motion of chiral N\'{e}el walls\cite{ryu2013chiral,emori2013dmi,torrejon2014interface,yang2015domain,torrejon2016tunable}. 

The pulse amplitude ($I$) dependence of the wall velocity ($v$) can be fitted using the following phenomenological equation\cite{thiaville2012dynamics,boulle2013domain,martinez2014current,torrejon2016tunable}
\begin{equation}
\label{DMI_saturation_eq}
v(I) = \pm
v_\mathrm{D} \left[ 1 + \left( \frac{I_\mathrm{D}}{I - I_\mathrm{C}}\right)^2 \right]^{-\frac{1}{2}},
\end{equation}
where $v_\mathrm{D} = \frac{\pi}{2} \gamma \Delta \mu_0 H_\mathrm{DM}$ is the saturation velocity, $\gamma$ is the gyromagnetic ratio and $\mu_0$ is the vacuum permeability. 
$\Delta = \sqrt{A_\mathrm{ex} / K_\mathrm{eff}}$ is the domain wall width, $A_\mathrm{ex}$ is the exchange stiffness, $H_\mathrm{DM} = D / (\mu_0 M_\mathrm{s} \Delta)$ is the DM exchange field and $D$ is the DM exchange constant. In Eq. (\ref{DMI_saturation_eq}), the saturation pulse amplitude $I_\mathrm{D}$ and the threshold amplitude $I_\mathrm{C}$ are phenomenological parameters. We fit the data shown in Figs. \ref{DMI_saturation}(a) and \ref{DMI_saturation}(b) with $v_\mathrm{D}$, $I_\mathrm{D}$ and $I_\mathrm{C}$ as independent fitting parameters: the solid lines show the fitting results. The fitting is in good agreement with the experimental results.
The DM exchange constant ($D$) extracted from the fitting is shown in Table \ref{para_table}. 
$D$ is obtained from the average $v_\mathrm{D}$ of up/down and down/up domain walls driven by positive and negative currents. $D$ is $\sim 30\%$ larger for sample B ($d_{\mathrm{Ta}} \sim 2.3$ nm) than sample A ($d_{\mathrm{Ta}} \sim 0.5$ nm). We infer that the difference in $D$ with respect to $d_{\mathrm{Ta}}$ may partly originate from structural\cite{hrabec2014dmi} and compositional differences\cite{loconte2015dmi} of the W/CoFeB interface (see Refs. \cite{soucaille2016dmi,gross2016dmi} for $D$ of similar heterostructures in the absence of current). 

\begin{figure}[t]
\includegraphics[scale=0.45]{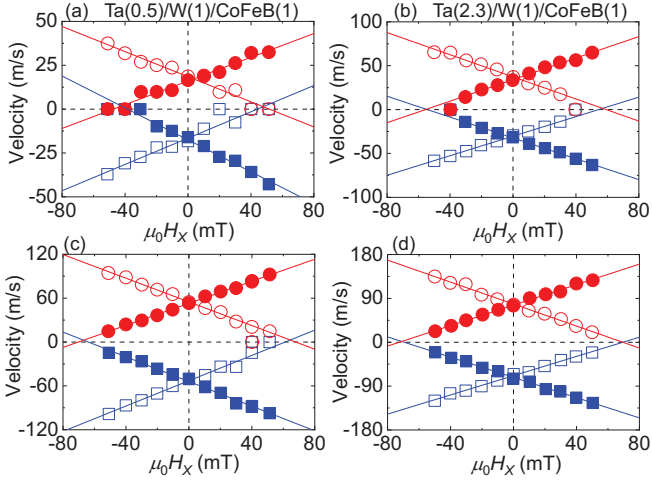}
	\caption{
	(a,b) $H_x$ dependence of the domain wall velocity. The pulse amplitude is (a) $\sim14$ V, (b) $\sim8$ V, (c) $\sim25$ V and (d) $\sim18$ V. The red circles (blue squares) show velocity when positive (negative) current is applied. The filled (open) symbols represent the velocity of up/down (down/up) walls. The solid lines show linear fit to the data. (a,c) sample A, (b,d) sample B. 
\label{VelHx}}
\end{figure}

To study the relation between the DM exchange constant and the current that flows across a domain wall, we have studied the in-plane field ($H_x$, field parallel to the $x$ axis) dependence of the domain wall velocity for the two samples. The results are shown in Fig. \ref{VelHx} for two different pulse amplitudes.
As evident, the velocity varies linearly with $H_x$ and can be fitted with a linear function\cite{emori2013dmi,ryu2013chiral,torrejon2014interface,ryu2014chiral,yang2015domain}. The field at which the velocity becomes zero provides information on the DM exchange field\cite{thiaville2012dynamics,martinez2014current}. The absolute values of the $x$-intercepts for the up/down and down/up walls driven by positive and negative currents are averaged to obtain the mean DM exchange field $H_\mathrm{DM}$. Taking the mean value of the four cases eliminates effects of $H_z$, if any, that arises due to slight misalignment of the sample with respect to the field axis. 

\begin{figure}[b]
\includegraphics[scale=0.4]{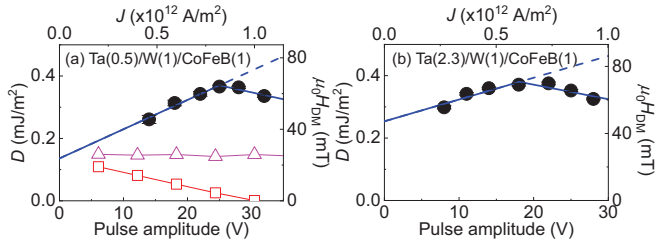}
	\caption{
	(a,b) Pulse amplitude dependence of the DM exchange constant $D$ for (a) sample A and (b) sample B. The top and right axes display the corresponding current density $J$ and $H_\mathrm{DM}$, respectively. The solid circles show the experimental data. The error bars represent the uncertainty of $H_{\textrm{DM}}$ due to the linear fitting of the velocity vs. $H_x$. The solid lines show linear fit to the data in appropriate current range. The open squares and triangles in (a) show $D$ vs. $J$ calculated using micromagnetic simulations. The adiabatic STT is turned off (on) for the calculations presented with the open triangles (squares). 
\label{DMI_voltage}}
\end{figure}

The DM exchange constant $D$ is estimated from $H_\mathrm{DM}$ using the relation $\mu_0 H_\mathrm{DM} = D / (M_\mathrm{s} \Delta)$ (we use $A_\mathrm{ex} \sim 1.5 \times 10^{-11}$ J/m (Refs. \cite{eyrich2014exchange,devolder2016exchange})).
The pulse amplitude dependence of $H_\mathrm{DM}$ and $D$ are shown in Figs. \ref{DMI_voltage}(a) and \ref{DMI_voltage}(b), solid circles, for the two samples.
The top axis shows the corresponding average current density $J$ that flows through W, CoFeB and the bottom Ta layers: here an uniform current flow is assumed. 
We find that $D$ depends on $J$. For both samples, $D$ increases with increasing $J$ for small current. The change in $D$ with $J$ is significant: the difference between the maximum and minimum $D$ is $\sim 30-40\%$. We also find that $D$ tends to decrease at larger current for both samples. 

\begin{figure}[b]
\includegraphics[scale=0.4]{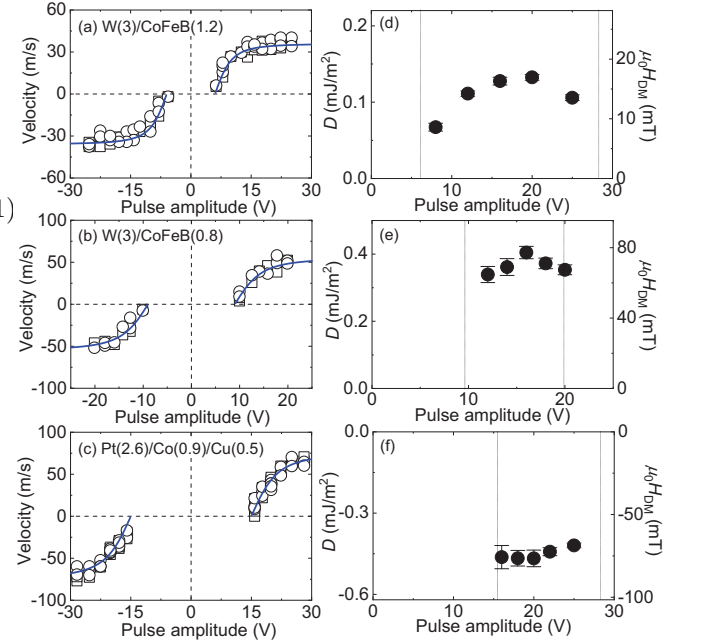}
	\caption{
	(a-c) Pulse amplitude dependence of domain wall velocity for samples C (a), E (b) and F (c). The circles (squares) show velocity of up/down (down/up) domain walls. The blue solid lines show the fitting results using Eq. (\ref{DMI_saturation_eq}). The fitting parameters are summarized in Table~\ref{fit_table} of the Supplementary material. (d-f) Pulse amplitude dependence of the DM exchange constant $D$ of samples C (d), E (e) and F (f). The right axis displays the corresponding $H_\mathrm{DM}$. The error bars represent the uncertainty of $H_{\textrm{DM}}$ due to the linear fitting of the velocity vs. $H_x$. The vertical dashed lines in (d-f) indicate the range of pulse amplitude used to measure the wall velocity presented in (a-c). 
\label{DMI_voltage2}}
\end{figure}

To clarify the origin of the current induced modulation of the DM exchange interaction, we have also studied the effect in other systems. Figure \ref{DMI_voltage2} shows the results from three different systems: sub./W(3)/Co$_{20}$Fe$_{60}$B$_{20}$($t_{\mathrm{FM}}$)/MgO(2)/Ta(1) with $t_{\mathrm{FM}} \sim 1.2$ nm (sample C) and $t_{\mathrm{FM}} \sim 0.8$ nm (sample E) and sub./Ta(2)/Pt(2.6)/Co(0.9)/Cu(0.5)/Pt(0.6)/MgO(2)/Ta(1) (sample F). Samples C and E are evaluated to study the influence of the FM layer thickness. Sample F consists of a Pt/Co/Pt based structure with a thin Cu layer inserted between the Co and top Pt layers to break the structural inversion symmetry.  

The pulse amplitude dependence of the wall velocity is plotted in Figs. \ref{DMI_voltage2}(a-c). For all samples, we find the wall velocity increases with increasing pulse amplitude until it saturates, and the domain wall moves along the current flow. Since the signs of the spin Hall angle of W and Pt are opposite\cite{emori2013dmi,torrejon2014interface} these results suggest that the magnetic chirality of the Pt/Co interface is opposite to that of W/CoFeB.
The pulse amplitude dependence of $D$ is presented in Figs. \ref{DMI_voltage2}(d-f).
Interestingly, we find a strong change in $D$ with the pulse amplitude for sample C, i.e. W/CoFeB with thicker FM layer. The current density dependence of $D$ is similar to that found in samples A and B. Although the trend is less clearer, $D$ varies with the current density for the thinner FM layer stack (i.e. sample E). In contrast, we find a nearly constant $D$ against the pulse amplitude for sample F, albeit the large change in the velocity with the pulse amplitude.

We have performed micromagnetic simulations to study the current density dependence of $D$. Under the influence of the damping-like torque that arises from the spin Hall induced spin current, the $H_x$ dependent wall velocity is calculated to obtain $H_{\textrm{DM}}$, from which $D$ is calculated. In addition, the effect of the adiabatic spin transfer torque (STT) is studied. (The non-adiabatic STT is neglected here since it was reported to be small in CoFeB\cite{fukami2011cfb}.) We model the case of sample A using parameters obtained from experiments (see Supplementary material for the details of the parameters used). The calculated $D$ is plotted as a function of current density in Fig. \ref{DMI_voltage}(a): the open squares and the open triangles represent $D$ when the adiabatic STT is turned on and off, respectively. Without the adiabatic STT, $D$ does not change with current. In contrast, $D$ tends to decrease with increasing current density when the adiabatic STT is turned on. Thus the drop in $D$ at larger current may be associated with the influence of the adiabatic STT.
Note that the field-like torque\cite{miron2010rashba,miron2011dw,kim2013torque,garello2013torque} that typically appears in HM/FM bilayers does not influence $H_{\textrm{DM}}$ according to the one dimensional model of a domain wall\cite{torrejon2014interface}.

\begin{figure}[b]
\includegraphics[scale=0.5]{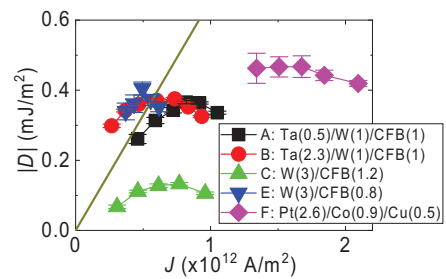}
	\caption{
	Current density $J$ dependence of DM exchange constant $D$ for samples A-C, E and F. The error bars represent the uncertainty of $H_{\textrm{DM}}$ due to the linear fitting of the velocity vs. $H_x$. The solid line show the relation $D = \frac{\hbar}{e} J$. 
\label{DMI_voltage3}}
\end{figure}

We thus infer that the current induced change of $D$ may arise from two competing effects: the adiabatic STT and a second effect that tends to increase $D$ with increasing current.
One of the possible sources of the latter is the spin Doppler effect. 
Model calculations indicate that the DM exchange interaction emerges from an equilibrium spin current\cite{kim2013rashbadmi,kikuchi2016dzyaloshinskii,freimuth2017dmi,tatara2019physE} and that $D$ can be modulated with external source of spin current\cite{tatara2019physE,freimuth2017dmi}.
In Fig. \ref{DMI_voltage3}, we plot $D$ against the current density $J$ for the samples studied and compare it with what the model predicts. 
(To estimate $J$, we assume uniform current flows through W and CoFeB layers for samples C and E. For sample F, an uniform current flow through the highly conductive Pt, Co and Cu layers is assumed.)
The model dictates that the change in $D$ with the spin current density $J_s$ is related by fundamental constants, i.e. $D = \frac{\hbar}{e} J_s$, where $\hbar$ and $e$ are the reduced Planck constant and the elementary charge, respectively. This relation is plotted by the solid line in Fig. \ref{DMI_voltage3} with $J_s \approx J$. Interestingly, we find many of the samples exhibit, at small $J$, an increase in $D$ with $J$ similar to what the model predicts: the rate of increase will become even closer if one takes into account the effect of the adiabatic STT. For sample F, larger $|J|$ is needed to move domain walls due to the stronger pinning and the smaller spin current from the Pt layer compared to that of the W layer. In this range of $|J|$, changes in $D$ against $J$ tends to be small, consistent with previous reports\cite{ryu2013chiral}.

The source of the spin current that causes the spin Doppler effect, however, remains as an issue.
In general, the energy density of the DM interaction can be expressed as:
\begin{equation}
\label{eq:DMEnergy}
\epsilon_{\textrm{DM}} = \sum_{i,a=x,y,z} D_i^a (\partial_{i} \bm{m} \times \bm{m} )_a,
\end{equation}
where $\bm{m}$ denotes the direction of local magnetic moments and $D_i^a$ represents a component of the DM vector.
The spin Doppler effect dictates that $D_i^a$ is equivalent to a spin current with polarization along $a$ and flow along $i$. 
Note that $D_i^a$ must be orthogonal to the local magnetic moment ($D_i^a \propto m^a$ does not contribute to $\epsilon_{\rm DM}$) to induce chiral magnetic structure.
Here we consider a quasi one dimensional (1D) system (long axis along $x$) in which the spatial profile of the magnetization changes only along the $x$ axis and the magnetic easy axis points along the $z$ axis.
In the spin Doppler model, this restriction suggests that an uniform spin (polarized) current that flows along $x$ predominantly contributes to the DM interaction.

In the HM/FM bilayer system under consideration, the current induced spin polarization via the Edelstein effect\cite{edelstein1990,miron2010rashba,miron2011dw,manchon2015review} and the spin polarized current\cite{zhang2004beta} that flows within the FM layer include flow of electrons along $x$ with polarization transverse to the local magnetic moments.
It turns out that, in first order approximation, both effects do not contribute to the modification of $D$ (see Supplementary material for the details).
The spin Hall effect of the HM layer does not, in general, generate a spin current that flows along $x$ (the flow direction is along $z$).
However, the spin accumulation at the HM/FM interface induced by the spin Hall effect can be driven by the current that flows within the FM layer and be the source of $D_x^y$. The sign of spin (polarized) current generated by such process will not change upon reversing the current flow: thus one expects flow direction independent changes in $D$, in agreement with the experiments.
Other possibilities include spin current generated at the interface\cite{wang2016interface,amin2018interface,amin2019fmspin} or dynamical effects that only pertains to moving domain walls\cite{freimuth2018dmi}. Further investigation is required to identify the cause of the current induced modulation of $D$\cite{karnad2018dmi}.

In summary, we have studied the interfacial Dzyaloshinskii-Moriya interaction against current in W/CoFeB and Pt/Co heterostructures. In contrast to the Pt/Co based heterostructures in which a nearly current independent $D$ is found, the DM exchange constant of the W/CoFeB based heterostructures increases linearly with current at lower current and tends to decrease at larger current. The current flow direction has little impact on the modulation of the DM exchange constant $D$. 
The rate of increase of $D$ with the current density at lower current is in agreement with the model based on the spin Doppler effect.
According to micromagnetic simulations, the drop in $D$ at larger current may be associated with the adiabatic spin transfer torque. These results indicate that the DM interaction at the heavy metal/ferromagnetic metal interface originates, if not entirely, from the exchange of equilibrium spin current at the interface. Our findings thus suggest that DM interaction is not a given material parameter of each interface but can be controlled externally using current. Such external control of DM interaction can significantly expand the scope of research on chiral magnetism in thin film heterostructures. 


\begin{acknowledgments}
Acknowledgments: We thank G. Tatara, T. Koretsune, H. Kohno, Y. Imai and S. Takahashi for fruitful discussions. This work was partly supported by JSPS Grant-in-Aid for Specially Promoted Research (15H05702), Scientific Research (16H03853), Casio Foundation, and the Center of Spintronics Research Network of Japan. Y.-C.L. is an International Research Fellow of the JSPS. T. K. is a Yukawa Research Fellow supported by the Yukawa Memorial Foundation.
\end{acknowledgments}

\bibliography{reference_060719}

\clearpage
\section*{Supplementary material}
\section{Experimental results}
\subsection{Sample preparation}
For samples A and B (sub./Ta($d_{\mathrm{Ta}}$)/W(1)/Co$_{20}$Fe$_{60}$B$_{20}$(1)/MgO(2)/Ta(1) (thickness in nm)), the Ta underlayer thickness $d_{\mathrm{Ta}}$ is varied using a moving shutter during the deposition process to linearly change the Ta layer thickness across the substrate. 
Similarly, the CoFeB layer thickness $t_{\mathrm{FM}}$ is varied using a moving shutter for samples C and E (sub./W(3)/Co$_{20}$Fe$_{60}$B$_{20}$($t_{\mathrm{FM}}$)/MgO(2)/Ta(1)).

\subsection{Magnetic properties of the heterostructures}
\begin{figure}[h!]
	\includegraphics[scale=0.4]{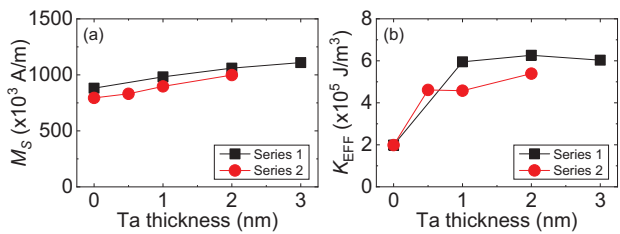}
	\caption{Magnetic properties of the sub./Ta($d_{\mathrm{Ta}}$)/W(1)/Co$_{20}$Fe$_{60}$B$_{20}$(1)/MgO(2)/Ta(1) heterostructures. (a) $M_\mathrm{s}$ and (b) $K_\mathrm{eff}$ as a function of the Ta layer thickness $d_{\mathrm{Ta}}$ for samples A and B. 
	\label{Ms_and_Keff}}
\end{figure}

The saturation magnetization $M_\mathrm{s}$ and the effective magnetic anisotropy $K_\mathrm{eff}$ are measured using vibrating sample magnetometry (VSM) for films made of sub./Ta($d_{\mathrm{Ta}}$)/W(1)/Co$_{20}$Fe$_{60}$B$_{20}$(1)/MgO(2)/Ta(1) ($d_{\mathrm{Ta}} \sim$0, 0.5, 1, 2, 3 nm). VSM is performed on individual films with constant film thickness across the substrate. The results are shown in Fig. \ref{Ms_and_Keff}. 
Two series of films are made at different times. Although the film structure and the deposition/annealing condition are set exactly the same, trivial differences in the deposition/annealing condition influence the magnetic anisotropy energy. Series 1 represents the magnetic properties of the samples presented in the main text; series 2 is made long after series 1 was deposited. Data from series 1 are linearly interpolated to obtain the corresponding $M_\mathrm{s}$ and $K_\mathrm{eff}$ at the Ta layer thicknesses for samples A and B. 

\subsection{Current pulse shape}
\begin{figure}[h!]
	\includegraphics[scale=0.45]{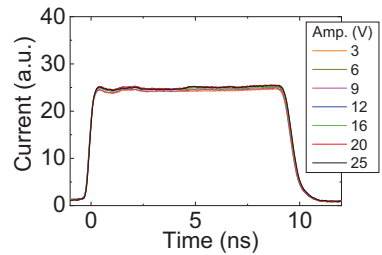}
	\caption{The measured current pulse that flows into the device. The measured pulse is normalized by the set pulse amplitude $A$ of the pulse generator.
	\label{currentpulse}}
\end{figure}

The pulse generator used in measuring the domain wall velocity has rise and fall times of $<$0.3 ns and $\sim$0.75 ns, respectively. The bandwidth of the GSG probe used to contact the device and the SMA cables are 40 GHz and 20 GHz, respectively.
Time domain reflectometry is used to characterize the shape of the current pulse that flows into the device. 
Since the device employs a short-end termination, we use a reflection geometry. 
See Ref. \cite{torrejon2016tunable} for the details of the measurement setup.
The film stack of the device evaluated is similar to sample C (the CoFeB thickness is $\sim$0.1 nm thinner than sample C); the wire geometry is identical. Figure \ref{currentpulse} shows the measured current pulse that flows into the wire. The pulse amplitude $A$ is varied and the pulse width is set to $\sim$9 ns. The measured current pulse is normalized by $A$ to illustrate the variation of the pulse shape, if any, with the pulse amplitude. As evident, the current pulse is close to a square pulse and the pulse shape (and consequently the pulse width) hardly changes with $A$.

\subsection{Fitting parameters of the velocity vs. pulse amplitude}
The pulse amplitude dependence of the domain wall velocity is fitted using Eq. (1). The fitting parameters are summarized in Table \ref{fit_table}. The fitting is performed without taking into account the current dependence of the DM exchange constant $D$. The fitting results are shown by the solid blue lines in the corresponding figures (Figs. 2(a,b) and 5(a-c)). 

\begin{table}[h]
\caption{Parameters used to fit the pulse amplitude dependence of the domain wall velocity\footnote{Equation (1) is used for the fitting. The DM exchange constant $D$ is assumed to be a constant.}.\label{fit_table}}
	\begin{tabular}{p{1.5cm} p{1.5cm} p{1.5cm} p{1.5cm} p{1.5cm}}
	\hline \hline
	sample &	$v_D$ & $I_D$ & $I_C$ & data\\
	\hline
	A  & 72 m/s & 15 V & 10 V & Fig. 2(a)\\
	B  & 84 m/s & 7 V & 5 V & Fig. 2(b)\\
	C  & 36 m/s & 5 V & 6 V & Fig. 5(a)\\
	E  & 55 m/s & 6 V & 9 V & Fig. 5(b)\\
	F  & 75 m/s & 5 V & 15 V & Fig. 5(c)\\
	\hline \hline
	\end{tabular}
\end{table}

\begin{figure}[h]
\includegraphics[scale=0.45]{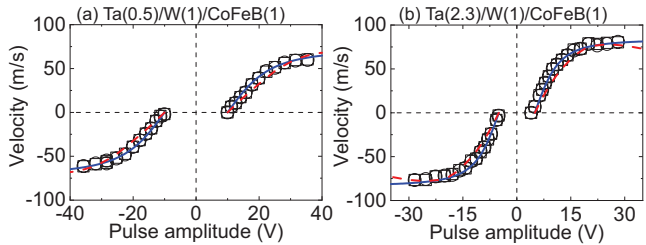}
	\caption{
	(a,b) Symbols and blue solid lines are reproduced from Fig. 2. Pulse amplitude dependence of domain wall velocity of (a) sample A and (b) sample B. The circles (squares) show velocity of up/down (down/up) domain walls. The blue solid lines show the fitting results using Eq. (1). The red dashed lines include the current dependent $D$ obtained from Fig. 4 when calculating the velocity with Eq. (1). The fitting parameters for the blue solid line (with current independent $D$) and the red dashed lines (with current dependent $D$) are summarized in Tables~\ref{fit_table} and \ref{fit_table2}, respectively.
\label{DMI_saturation_supp}} 
\end{figure}

Since we have used a constant $D$ to fit the pulse amplitude dependence of the velocity (Fig. 2, blue solid line), we recalculate the quantity using the results from Fig. 4.
The pulse amplitude dependent $D$ is fitted with two linear functions, as shown by the blue solid lines in Figs. 4(a) and 4(b).
We assume the phenomenological parameter $I_\mathrm{D}$ is linearly proportional to $D$\cite{thiaville2012dynamics}, i.e. $I_\mathrm{D} = a D$. The calculated velocity, with $a$ and $I_\mathrm{C}$ as adjustable parameters, is shown by the red dashed line in Figs. \ref{DMI_saturation_supp}(a) and \ref{DMI_saturation_supp}(b) for samples A and B, respectively. 
We find a relatively good agreement between the experimental results and the calculations. Table  \ref{fit_table2} displays the parameters used for this fitting. 

\begin{table}[h]
\caption{Parameters used to fit the pulse amplitude dependence of the domain wall velocity\footnote{Equation (1) is used for the fitting with the assumption $I_\mathrm{D} = a D$. The pulse amplitude dependence of $D$ is obtained from the two linear fittings shown in Figs. 4(a) and 4(b).} with current dependent DM exchange constant $D$. \label{fit_table2}}
	\begin{tabular}{p{1.5cm} p{3cm} p{1.5cm} p{2cm}}
	\hline \hline
	sample &	$a$ & $I_\mathrm{C}$ & data\\
	\hline
	A  & $8.6 \times 10^4$ V m$^2$/J & 10 V & Fig. 2(a)\\
	B  & $3.1 \times 10^4$ V m$^2$/J & 5 V & Fig. 2(b)\\
	\hline \hline
	\end{tabular}
\end{table}

\section{Micromagnetic simulations}
Micromagnetic simulations are performed using a home-made program.
The following equation is numerically solved to study the magnetization dynamics of a ferromagnetic metal (FM) layer placed next to a non-magnetic metal (NM) layer that generates spin current via the spin Hall effect.  
\begin{equation}
\begin{aligned}
\frac{\partial \bm{m}}{\partial t} = &- \gamma \bm{m} \times \big( \bm{H} + H_{\textrm{SH}} \bm{m} \times \bm{p} \big)\\
&+ \alpha \bm{m} \times \frac{\partial \bm{m}}{\partial t} - u (\bm{j}_{\textrm{FM}} \cdot \nabla) \bm{m}
\label{eq:LLG}
\end{aligned}
\end{equation}
where $t$ is time, $\gamma$ is the gyromagnetic ratio, $\bm{m}$ represents the magnetization unit vector and $\alpha$ is the Gilbert damping constant of the FM layer. The current induced effective field caused by the spin current that diffuses in from the NM layer is defined as $H_{\textrm{SH}} \equiv \frac{\hbar \theta_{\textrm{SH}} J_{\textrm{NM}}}{2 e M_\textrm{s} t_{\textrm{FM}}}$, where $\hbar$, and $e$ are the reduced Planck constant and the elementary charge, respectively, $\theta_{\textrm{SH}}$ and $J_{\textrm{NM}}$ are the spin Hall angle of and the current density that flows in the NM layer, respectively, $M_\textrm{s}$ and $t_{\textrm{FM}}$ are the saturation magnetization and the thickness of the FM layer, respectively. $\bm{p}$ is the average spin polarization of the electrons impinging on the FM layer from the NM layer. The adiabatic spin torque term $u$ is defined as $u=\frac{\mu_{\textrm{B}} P J_{\textrm{FM}}}{e M_\textrm{s}}$, where $\mu_{\textrm{B}}$ is the Bohr magneton, $P$ and $j_{\textrm{FM}}$ are the spin polarization of and the unit current density that flows in the FM layer, respectively.

The effective magnetic field is calculated from the magnetic energy density using the relation $\bm{H} = -\frac{1}{M_\textrm{s}}\frac{\delta \epsilon}{\delta \bm{m}}$, where $\delta$ represents a functional derivative. The energy density takes the following form in the system under consideration (the film plane normal is along $z$):
 \begin{equation}
\begin{aligned}
\epsilon &= A_\textrm{ex} (\nabla \bm{m})^2 + K_\textrm{u} (1- m_z^2) - M_\textrm{s} \bm{m} \cdot \bm{H}_\textrm{ext}\\
&+ D \Big[ \big(m_x \frac{\partial m_z}{\partial x} - m_z \frac{\partial m_x}{\partial x} \big) + \big(m_y \frac{\partial m_z}{\partial y} - m_z \frac{\partial m_y}{\partial y} \big) \Big]\\ 
&- \frac{1}{2} M_\textrm{s} \bm{m} \cdot \bm{H}_\textrm{dem}
\label{eq:LLGEnergy}
\end{aligned}
\end{equation}
$A_\textrm{ex}$, $K_\textrm{u}$, and $D$ are the exchange constant, uniaxial effective perpendicular magnetic anisotropy energy density, and the Dzyaloshinskii-Moriya (DM) exchange constant, respectively, of the FM layer. The DM exchange interaction is assumed to act on the magnetic moments placed at the interface. $\bm{H}_\textrm{ext}$ and $\bm{H}_\textrm{dem}$ are the external and demagnetizing magnetic fields. $\bm{H}_\textrm{dem}$ is calculated numerically.

A nanowire made of a FM layer is divided into identical rectangular prisms (cells) with dimensions of $2.5 \times 2.5 \times 1.0$ nm$^3$.
The width and the thickness of the nanowire are 320 nm and 1 nm, respectively. 
To mimic the random pinning of domain walls, dispersion of uniaxial anisotropy is included: the standard deviation of the uniaxial anisotropy around its mean value is defined in the simulation.
A moving boundary condition is employed, in which the calculation frame always places the domain wall in the center.
The domain wall velocity is calculated from the distance the domain wall travels during the pulse application.

The following parameters are used in the simulations to model sample A. 

\begin{table}[h!]
\caption{Parameters used in micromagnetic simulations.\label{umag_table}}
	\begin{tabular}{p{6cm} p{2cm}}
	\hline \hline
	parameter &	value\\
	\hline
	Gilbert damping constant $\alpha$	&	0.1\\
	Exchange stiffness $A_\mathrm{ex}$	&	$1.5 \times 10^{-11}$ J/m\\
	Dzyaloshinskii-Moriya interaction $D$	&	0.14 mJ/m$^2$\\
	Spin Hall angle $\theta_\mathrm{SH}$	&	-0.2\\
	Saturation magnetization $M_\mathrm{s}$	&	930 kA/m\\
	Magnetic anisotropy energy\footnote{$K_\mathrm{eff}$ = $K_\mathrm{u} - \frac{1}{2} \mu_0 M_\textrm{s}^2$} $K_\mathrm{eff}$	&	$4.0 \times 10^5$ J/m$^3$\\
	Standard deviation of $K_\mathrm{eff}$	&	10\%\\
	Spin polarization $P$	&	0.7\\
	Pulse width	&	9 ns\\
	\hline \hline
	\end{tabular}
\end{table}

\section{The spin Doppler effect}
The Hamiltonian of the system with the DM interaction reads
\begin{equation}
\begin{aligned}
\label{eq:Hamiltonian}
\mathcal{H}_{\textrm{DM}} &= \int d^3r D^a_i (\bm{\nabla}_i \bm{m} \times \bm{m} )_a \\
&= \int d^3r \epsilon_{\textrm{DM}} \big( \bm{m}(\bm{r}), \nabla \bm{m}(\bm{r})\big),\\
\epsilon_{\textrm{DM}} \big(& \bm{m}(\bm{r}), \nabla \bm{m}(\bm{r}) \big) \equiv D^a_i (\bm{\nabla}_i \bm{m} \times \bm{m} )_a.
\end{aligned}
\end{equation}
The effective field $\bm{H}$ takes the form of
\begin{equation}
\label{eq:Field}
H_i = - \frac{1}{M_S} \frac{\delta \epsilon_{\textrm{DM}}}{\delta m_i} = -\frac{1}{M_S} \big( \frac{\partial \epsilon_{\textrm{DM}}}{\partial m_i} - \partial_j \cdot \frac{\partial \epsilon_{\textrm{DM}}}{\partial (\partial_j m_i)} +... \big) ,
\end{equation}
where $\displaystyle{\frac{\delta \epsilon_{\textrm{DM}}}{\delta m_i}}$ represents a functional derivative and the indices $i, j=1,2,3$ (or $x,y,z$). 
Repeated indices imply summation.
The spin Doppler effect dictates that $D_i^a$ is equivalent to a spin current with polarization along $a$ and flow along $i$. 
Note that $D_i^a$ is transverse to the local magnetic moment.

As an example, with a Rashba type spin orbit coupling:
\begin{equation}
\begin{aligned}
\label{eq:Rashba}
D= D_0
\begin{bmatrix}
0 &1 &0\\
-1 &0 &0\\
0 &0 &0
\end{bmatrix}.
\end{aligned}
\end{equation}
Substituting Eq. (\ref{eq:Rashba}) (i.e. $D_x^y = D_0$, $D_y^x = -D_0$) into Eq. (\ref{eq:Hamiltonian}) and taking the functional derivative using Eq. (\ref{eq:Field}) returns
\begin{equation}
\begin{aligned}
\label{eq:Field_Rashba}
\bm{H} = \frac{2 D_0}{M_S} 
\begin{bmatrix}
-\partial_x m_z, &-\partial_y m_z, &\partial_x m_x + \partial_y m_y
\end{bmatrix}.
\end{aligned}
\end{equation}
This agrees with the common form used to describe the DM interaction in HM/FM bilayer systems (see e.g. Eq. (7) of Ref. \cite{rohart2013skyrmion}). 


\subsection{Current induced spin polarization}
The Edelstein effect\cite{edelstein1990} can generate spin current (accompanied with a charge current) that flows along $x$. 
However, simple model calculations show that the amount of spin current created by the charge current is compensated by the reduction in spin current with opposite spin moving against the former, both of which contribute to the spin Doppler effect in the same way.
Thus, in a first order approximation, the current induced spin accumulation due to the Edelstein effect does not influence $D$ via the spin Doppler effect.

\subsection{Spin polarized current in ferromagnets}
The spin polarized current that flows within the FM layer includes flow of electrons along $x$.
When current is passed along ferromagnets with non-uniform magnetic textures, it is known that a non-zero misalignment occurs between the conduction electron spin and the localized magnetic moments.
Such misalignment is the origin of the so-called non-adiabatic spin torque\cite{zhang2004beta} that can drive domain walls against the current flow.
With the misalignment, the conduction electrons that impinge on a local magnetic moment possess spin angular momentum that is transverse to the magnetic moment. 
Since such transverse component of the spin polarized current can possibly be the source of the spin Doppler effect, we examine its effect below.

We define $\bm{m}(x)$ and $\bm{\sigma}(x)$ as unit vectors of magnetization and the conduction electron spin, respectively, at position $x$ (time dependence is neglected here). The component of $\bm{\sigma}(x)$ transverse to $\bm{m}(x)$ is defined as $\bm{\bar{\sigma}} (x)$, i.e.
\begin{equation}
\begin{aligned}
\label{eq:sigmaperp}
\bm{\bar{\sigma}} (x) \equiv \bm{\sigma}(x) - \big(\bm{\sigma}(x) \cdot \bm{m}(x) \big) \bm{m}(x)
\end{aligned}
\end{equation}
We assume the conduction electrons move at a velocity $\bm{v}=v \hat{\bm{x}}$ due to the application of electric field. The current flow direction is defined as $u \equiv \frac{\bm{v}}{|\bm{v}|} \cdot \hat{\bm{x}}$. Under the application of electric field, the conduction electron spin carries information of the magnetization direction of the atom that it had passed by and is $\delta$ away from its current position, i.e.  
\begin{equation}
\begin{aligned}
\label{eq:misalignment}
\bm{\sigma}(x) = \bm{m}(x - u \delta) \approx \bm{m}(x) - u \delta \big( \partial_x \bm{m} \big)
\end{aligned}
\end{equation}
where we have assumed $\delta / \Delta \ll 1$ and $\Delta \equiv \sqrt{\frac{A_{\textrm{ex}}}{K_{\textrm{eff}}}}$ is the effective domain wall width.
Substituting Eq. (\ref{eq:misalignment}) into Eq. (\ref{eq:sigmaperp}) gives
\begin{equation}
\begin{aligned}
\label{eq:sigmaperp2}
\bm{\bar{\sigma}} (x) \approx - u \delta \big( \partial_x \bm{m} \big)
\end{aligned}
\end{equation}
The spin polarized current density $Q_x^a$ transverse to the local magnetic moment $\bm{m}$ is expressed as 
\begin{equation}
\label{eq:Qperp}
Q_x^a \propto \bar{\sigma}_a (x) v_x = - |v| \delta \big( \partial_x m_a \big)
\end{equation}
Substituting $Q_x^a$ into Eq. (\ref{eq:Hamiltonian}) as $D_x^a$ and calculating the effective field using Eq. (\ref{eq:Field}) returns $\bm{H} = 0$, which shows that the transverse spin polarized current that flows in the FM layer, with non-zero misalignment between the conduction electron spins and the local magnetic moments, does not contribute to the modification of $D$.

Note that the adiabatic STT and the non-adiabatic STT can cause changes in $H_{\textrm{DM}}$ via their respective torque on the domain walls, which results in a change of $D$ via the relation $\mu_0 H_\mathrm{DM} = D / (M_\mathrm{s} \Delta)$, see Fig. 4(a). The above discussion shows that the STT does not directly influence $D$ via the spin Doppler effect.



\end{document}